\documentclass[preprint,showpacs,preprintnumbers,amsmath,amssymb]{revtex4}

\usepackage[dvips]{graphicx}

\begin{document}
\DeclareGraphicsExtensions{.jpg,.png}
\title{Chaotic itinerancy, temporal segmentation and spatio-temporal
combinatorial codes}

\author{Juliana R. Dias}
\email{juliana@neuron.ffclrp.usp.br}
\author{Rodrigo F. Oliveira}
\email{rodrigo@neuron.ffclrp.usp.br}
\author{Osame Kinouchi}
\email{osame@ffclrp.usp.br}
\affiliation{Faculdade de Filosofia, Ci\^encias e
Letras de Ribeir\~ao Preto, Universidade de S\~ao Paulo, Av. dos
Bandeirantes 3900, 14040-901, Ribeir\~ao Preto, SP, Brazil}

\begin{abstract}

We study a deterministic dynamics with two time scales in a
continuous state attractor network. To the usual (fast) relaxation
dynamics towards point attractors (``patterns'') we add a slow
coupling dynamics that makes the visited patterns to loose
stability leading to an itinerant behavior in the form of punctuated
equilibria. One finds that the transition frequency
matrix between patterns shows non-trivial 
statistical properties in the chaotic itinerant regime. 
We show that mixture input patterns can be temporally segmented by the itinerant
dynamics. The viability of a combinatorial spatio-temporal
neural code is also demonstrated.

\end{abstract}


\maketitle

Several complex systems present a non-uniform rate of change, where stationary
states (``patterns'') suddenly loose their stability and are substituted by new
ones. Such punctuated behavior has been observed in a wide range of time scales
from evolutionary, economic, social and weather dynamics,
to brain behavior and laboratory devices such as laser cavities
\cite{Newman1985,Drossel,Kaneko2003}. Usually this ``itinerancy'' between states 
is thought as ``thermal'' transitions between deep valleys in
a rugged landscape, possibly in the glassy dynamics regime. Such
process, by definition, is stochastic so that times of transition
and the choice of the next pattern are random.
In this work, we consider the opposite spectrum of systems where
the loss of the patterns stability is
due to internal deterministic mechanisms \cite{Kaneko,Tsuda,Nara,Okuda}.
Of course, natural systems certainly falls between these two descriptions.

\begin{figure}[!tb]
\begin{center}
\includegraphics[width=0.8\columnwidth]{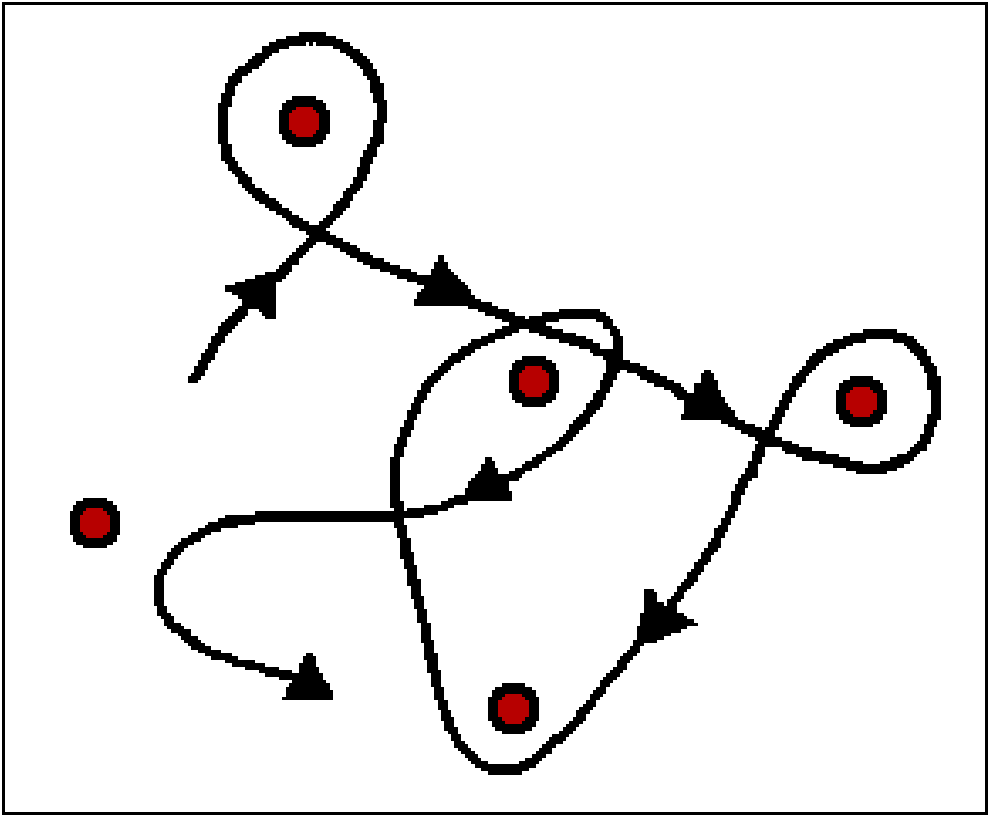}
\caption{\label{fig1} Schematic phase space showing an itinerant 
walk between point attractors.}
\end{center}
\end{figure}

The specific model studied here is a multistable system where
the relevant transitions occur when a stationary state (a point attractor)
looses stability so that the system falls into a new point attractor, and so on,
forming an itinerant trajectory (see Fig.~\ref{fig1}). The more general case of itinerancy
between several kinds of attractors (limit cycles, torus and low dimensional chaotic
attractors) has also been studied (see, for example, the special volume \cite{Kaneko2003}).

Our model consists of a
continuous state attractor network \cite{Marcus} storing $P$ patterns with an
added slow anti-Hebbian dynamics \cite{vanHemmen1997,Anderson1985,Hoshino}
(which may represent some coupling self-regulation by negative feedback).
The model has a discrete time parallel dynamics with a full
connected network, that is, it is a mean field coupled maps model:
\begin{equation}
S_i(t+1) = \mbox{tanh}[\gamma (h_i(t)+I_i(t))] \:, \label{fast}
\end{equation}
where $S_i$, the state of neuron $i = 1,\ldots, N$, is a real
variable in the interval $[-1,1]$, $h_i$ is the local field
$h_i(t) =   \sum_{j=1}^{N} J_{ij} S_j(t)$,
and $I_i(t)$ is an (eventual) external input.
The parameter $\gamma$ is the transfer function gain (in this paper,
$\gamma = 10$). Notice that we have called the units ``neurons'' only by
convention, since they could be better interpreted as
neural populations or basic units in a network (like glomeruli in the
olfactory system, species in ecological systems, population of agents in social systems etc).
With this interpretations, the mean field character present in the model is more plausible.

Eq.~(\ref{fast}) defines the dynamics for the 
fast variables $S_i(t)$
given the coupling matrix $J_{ij}$. In our model,
this matrix is slowly time dependent:
\begin{equation}
J_{ij}(t) = J_{ij}^H + J_{ij}^A(t)\:,
\end{equation}
where there is a constant Hebbian (``correlational'')
component $J_{ij}^H$ that stores $P$ patterns, defining
a basic attractor landscape, and a time dependent
anti-Hebbian part $J_{ij}^A(t)$ that modulates
this landscape and produces the escape events.

The Hebbian component has the usual form:
\begin{equation}
\label{Eq0}
J_{ij}^H = \frac{1}{N} \sum_\mu^P \xi_i^\mu \xi_j^\mu\:,
\end{equation}
where $\mathbf{\xi}^\mu = \{\xi_i^\mu\}, i = 1,\ldots,N; \mu = 1,\ldots, P$ are random
patterns to be stored. For convenience, we use binary random variables $\xi_i^\mu = \pm 1$.
As usual, we set $J_{ii} = 0$.

The present state $\mathbf{S}(t)$ of the system
modulates the ``energy'' landscape (defined by slow variables $J_{ij}(t)$), 
so that if the system is visiting a local minimum, that minimum slowly looses its
stability until turning out a saddle or a maximum and an escape
event occur. The change in the attractor landscape, however, is transient, having only an
exponentially decreasing memory of the past visited states.
So, the anti-Hebbian component has the form
\begin{equation}
\label{Eq1} J_{ij}^A(t+1) = (1 - \frac{1}{\tau}) J_{ij}^A(t) -
\frac{\epsilon}{N} S_i(t) S_j(t) \:.
\end{equation}
The initial condition is $J_{ij}^A(0) = 0$. The first term (the ``coupling memory decay'')
guarantees that any change produced by visiting some state
vanishes with characteristic time $\tau$ after the escape from that
state. The second term is the anti-Hebbian contribution,
parametrized by a step size $\epsilon$ and scaled by $1/N$ to
preserve compatibility with Eq.~(\ref{Eq0}).
So, the transition rate between patterns (or even the
possibility of such transitions) depends on the two
parameters $\tau$ and $\epsilon$.

The matrix $J_{ij}^H$ defines a permanent landscape of attractor basins
that is reversibly modulated but not destroyed by the anti-Hebbian term.
We notice that a similar dynamics has been studied by Kawamoto and Anderson
for the particular case $P=2$ intending to model visual pattern reversion
in the Necker cube \cite{Anderson1985}. Here we extend that study
to general number of patterns $P$.
Also Hoshino et al. \cite{Hoshino} used a similar anti-Hebbian dynamics
with an asymmetric coupling matrix to study transitions
between fixed points and cycles. Here we are interested in the chaotic itinerancy phase
that already appears with the simpler Hebbian matrix.

\begin{figure}[!tb]
\begin{center}
\includegraphics[width=\columnwidth]{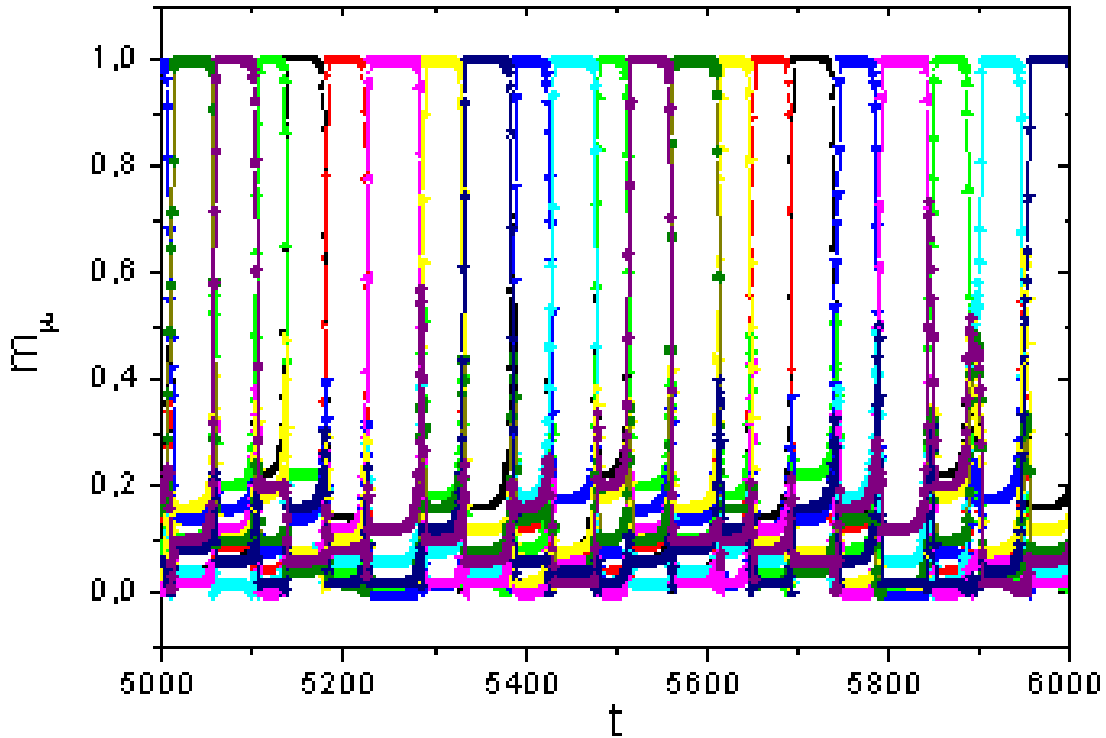}
\caption{\label{fig2} Time series of overlaps moduli $|m^\mu|$ for $P = 10$,
with $\epsilon = 0.009$, $\tau = 600$ and $N = 100$.}
\end{center}

\end{figure}

\begin{figure}[!tb]

\begin{center}
\includegraphics[width=\columnwidth]{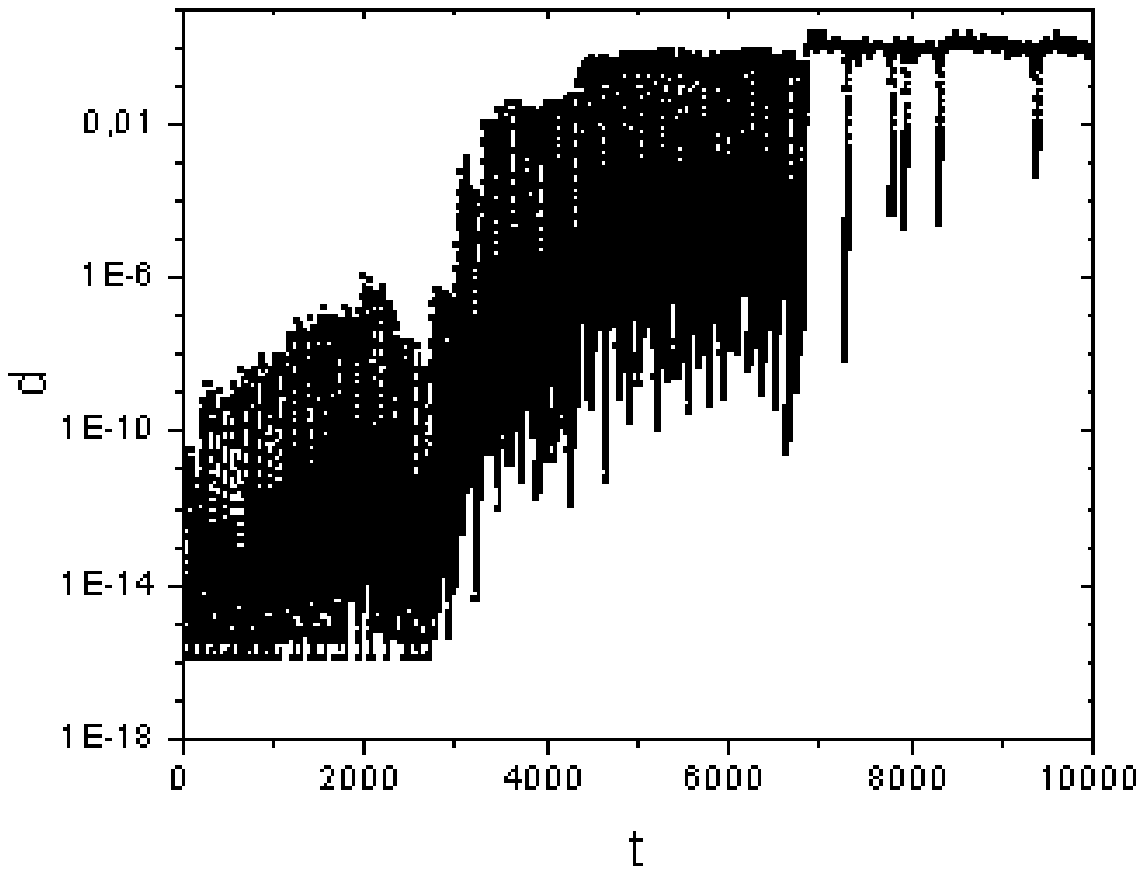}
\caption{\label{fig3}Distance between trajectories $d(t)$ as a function of time.
The two systems have identical initial coupling matrixes and differ only in the first 
neuron, $|S_1(0)-S_1^\prime(0)|=10^{-15}$.}
\end{center}

\end{figure}

Our results are presented in terms of low dimensional order parameters (``overlaps'')
that measure the correlation (cosine) between the state vector of the network
and the stored patterns, 
$m_\mu(t) \equiv
\sum_{j=1}^{N}S_j(t)\xi_j^\mu/(|\mathbf{S(t)}||\mathbf{\xi^\mu}|)$,
where  $|\mathbf{\xi^\mu}|= \sqrt{N}$ and $|\mathbf{S(t)}|$ are the vector
Euclidian norms.
In Fig.~\ref{fig2} we show an example of time series of the overlaps
for $P=10$ patterns. We define an (arbitrary but not crucial)
threshold $\lambda = 0.8$ so that we consider that a pattern
(or its anti-pattern) is being visited if $|m^\mu|  > \lambda$.
The $S(t)$ trajectory is indeed chaotic, as can be verified by observing
the distance $d(t)= \sum_i^N (S_i(t)- S_i^\prime(t))^2 $ 
between two orbits $\mathbf{S}(t)$ and 
$\mathbf{S}^\prime(t)$ with very small differences in initial conditions (Fig.~\ref{fig3}).

\begin{figure}[!tb]
\begin{center}
\includegraphics[width=\columnwidth]{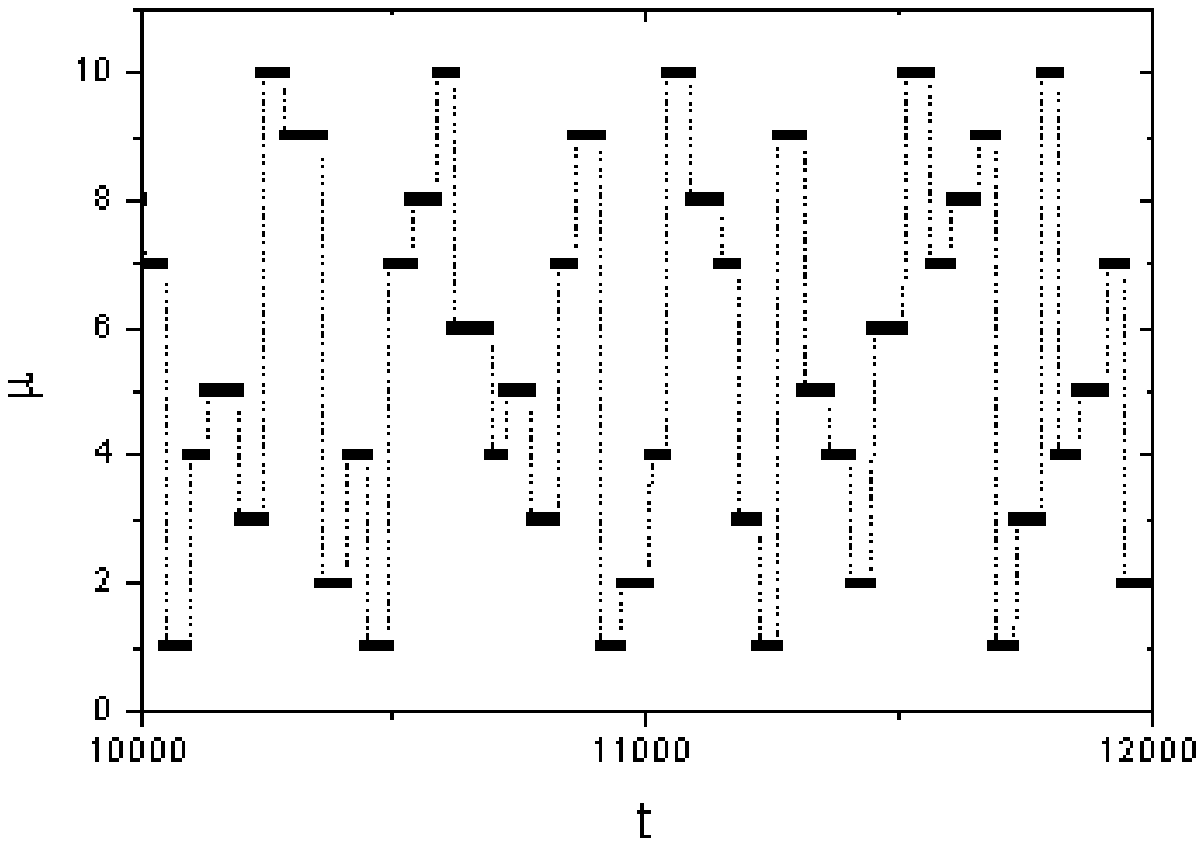}
\caption{\label{fig4} Label of the visited pattern as a function of time. The system
itinerates chaotically but with almost equal frequency between the stored patterns ($P=10$
and $N=100$).}
\end{center}
\end{figure}

From the time series one gets 
the symbolic time series where only appears the pattern (if any)
being visited (Fig.~\ref{fig4}). Although we report here only simulations with $N=100$ and 
 $\alpha=P/N=0.1$, we have checked that the itinerancy occurs until a effective
 critical storage $\alpha_c \propto 0.2$. This value is higher than the standard critical
 capacity $\alpha_c=0.14$ due to the effect of the anti-Hebbian term which is similar to 
 unlearning algorithms \cite{vanHemmen1997}. 

\begin{figure}[!tb]
\begin{center}
\includegraphics[width=\columnwidth]{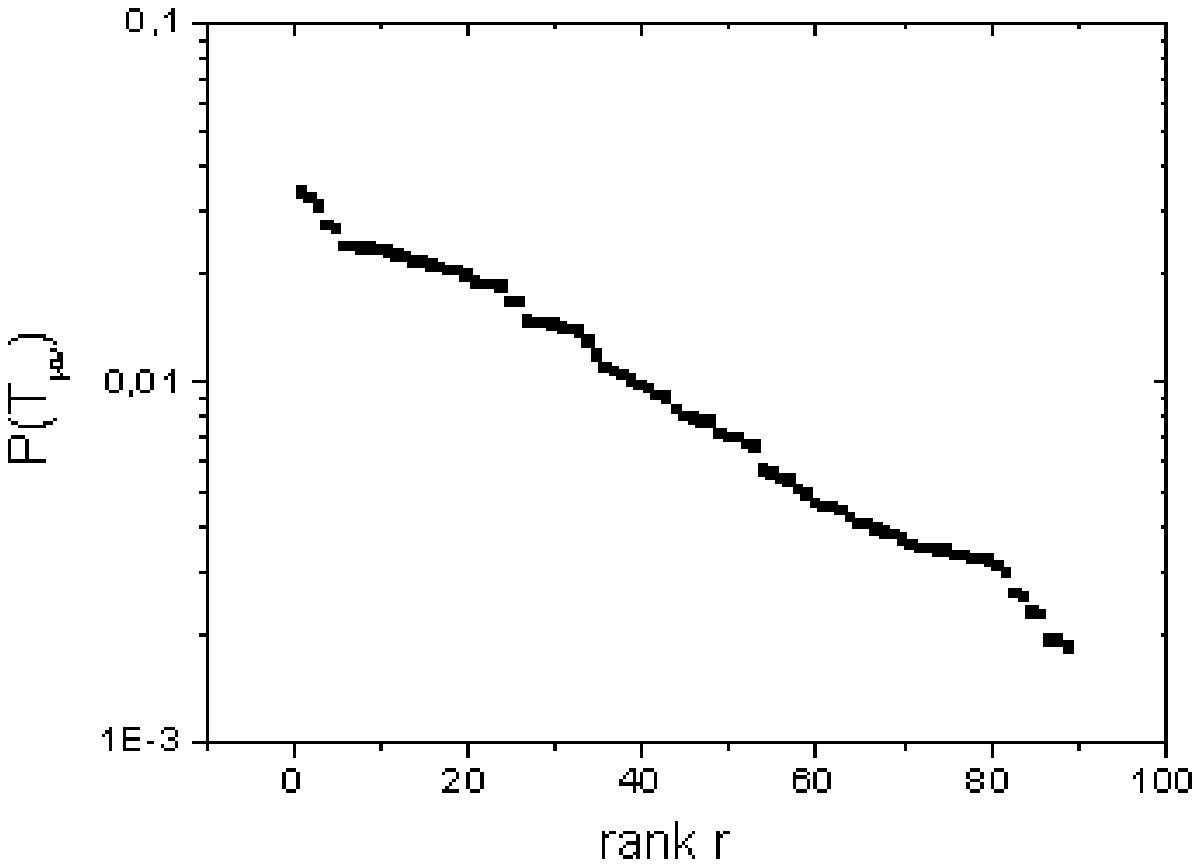}
\caption{\label{fig5} Rank plot of the values of the transition frequencies 
$T_{\mu\nu}$. Inset: Average residence times near each pattern.}
\end{center}

\end{figure}

From extensive symbolic time series one obtains
statistics about transitions and residence times  (Fig~\ref{fig5}). 
A not so obvious result is that, in a large parameter region,
all patterns are visited with almost equal frequency (Fig.~\ref{fig5} inset).
This finding is similar to that found in Nara model \cite{Nara}. 

We define $T_{\mu\nu}$ as the relative frequency matrix of transitions from pattern
$\nu$ to pattern $\mu$. In simple stochastic trap models \cite{Bouchaud, Martinez2004} 
the $T_{\mu\nu}$ elements are similar (there is no preferential transitions between
attractors). In our deterministic model, in contrast,
the transition matrix is very inhomogeneous, as can be seem by a
rank plot of  $T_{\mu\nu}$ values (Fig.~5).

Up to now we have reported the presence of chaotic itinerancy between
the stored patterns without external input ($I_i=0$). This
corresponds to a spontaneous itinerant network activity, as devised by Freeman
\cite{Freeman}. This author postulates that, under the presence of some
input, the chaotic itinerancy should collapse to an attractor of lower
dimension representing that input. Our model is a computational
implementation of Freeman's ideas. A new feature is that this lower
dimensional attractor also has a chaotic itinerant nature (with less components). 

Think about a mouse receiving a complex mixture of odours (each odour being represented
by an initial pattern in its olfactory glomeruli layer). Galan \em{et al.}\em{} have recently observed 
experimentally that glomerulli show Hebbian (correlational) plasticity, and that a Hopfield
network is a viable model of the glomerular layer \cite{Galan}.
Following Adachi et al.\cite{AAC}, we propose that chaotic
itinerancy between input components may be a mechanism to analyse such complex input by using
temporal segmentation. In a loose sense, we have a chaotic itinerancy 
projected into the subspace spanned by the patterns that compose the mixture input vector
\cite{AAC,Matykiewicz}. 

\begin{figure}[!tb]
\begin{center}
\includegraphics[width=\columnwidth]{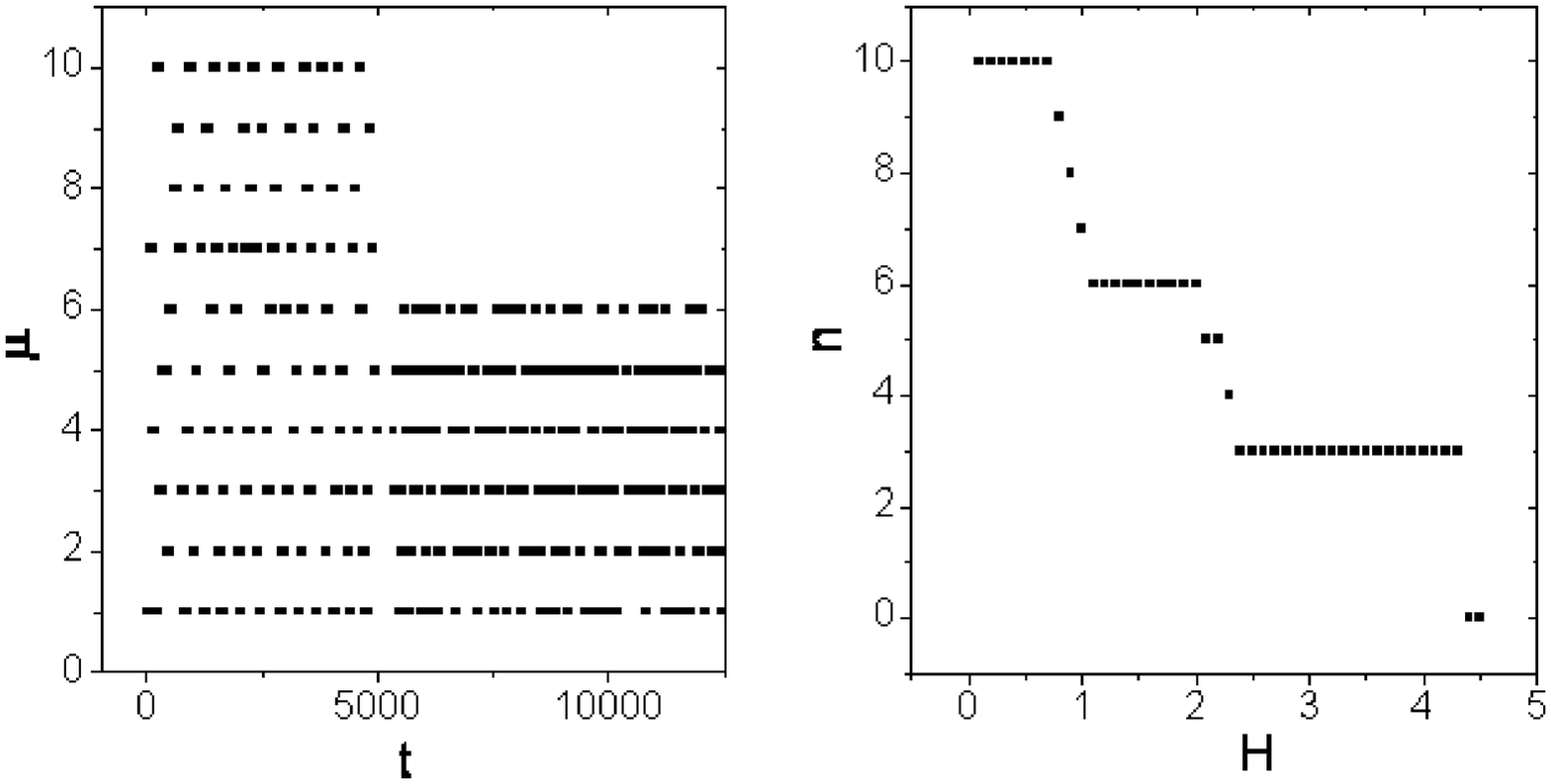}
\end{center}
\caption{\label{fig6} a) Network response to an input composed by $Q=6$ patterns ($\mu=1$
to $\mu=6$) with $H=1.2$. The input starts at $t=5000$. b) Total number of patterns visited
as a function of the input intensity $H$. Notice that for $n \le 6$, all patterns are
input components.}

\end{figure}

In Fig.~\ref{fig6}a we show the network response to inputs made of
a mixture of $Q$ patterns $
I_i= (H/\sqrt{Q}) \sum_{\nu=1}^{Q} \xi_i^\nu$,
where the input intensity is $H$. In Fig.~\ref{fig6}b we have
a bifurcation plot as a function of $H$ showing that there is a
robust interval where the system itinerates only between the $Q$
vectors that form the total input. This result is generic in parameter space and
is valid from $Q=1$ up to $Q=0.6P$. 
This means that the system makes a temporal segmentation
even if the input is composed by an extensive number of patterns.
For increasing odor intensity $H$ the nature of itinerancy 
changes following a series of plateaus. We conjecture that this could be the
theoretical correlate of a well known phenomenon where smells change 
abruptly of subjective character as a function of odour concentration \cite{smell}.

Although temporal segmentation is an intersting feature of the model, the more recent view is
that odours are represented as sequences of patterns of glomerular activation,
that is, a spatio temporal combinatorial code \cite{Laurent,Cleland}. The model allows such interpretation, if we
think the patterns $\xi^\mu$ as being not odour representations but as forming a basis to
combinatorially represent odours. In this scenario, we have upt to $C=P!/(Q!(P-N)!)$ possible
representations with $Q$ components. For example, with a typical vertebrate
glomerular layer with $N=1000$ glomeruli and assuming $P= 50$ and $Q=3$ (that is,
odours will be represented as sequences of three patterns), we have
$C= \approx 20,000 $ odours, which is already above the conjectured number of odours recognized by
mammals.

\begin{figure}[!tb]
\begin{center}
\includegraphics[width=0.7\columnwidth]{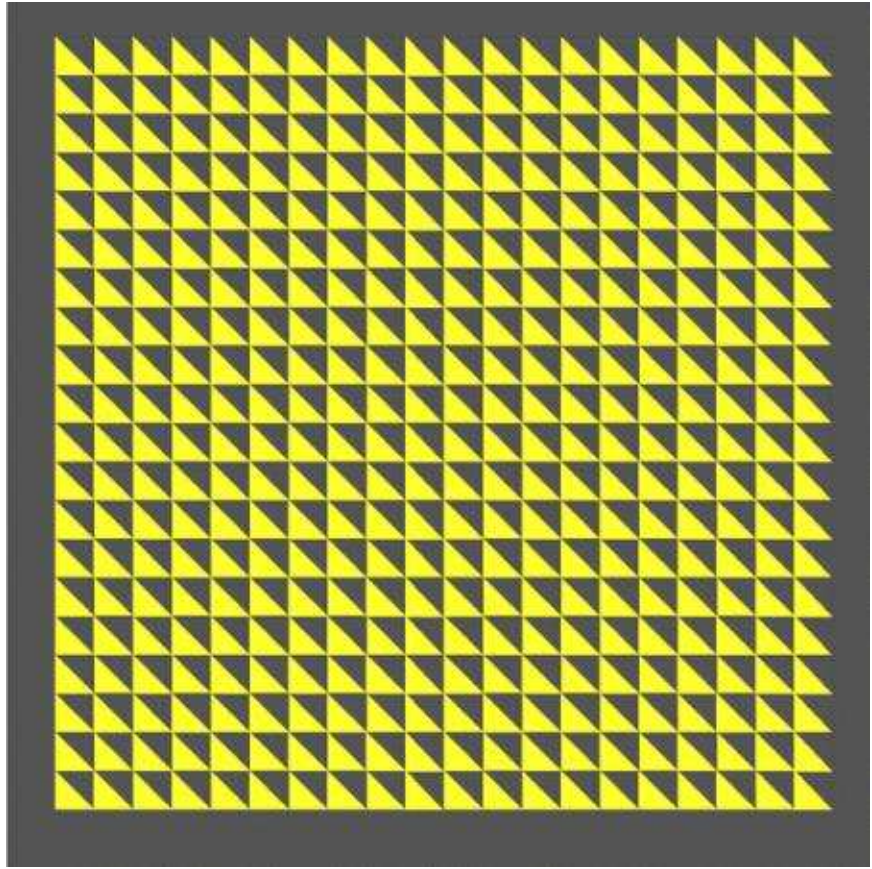}
\end{center}
\caption{\label{fig7} Example of visual multistability \cite{Multifigure}
where transitions between several patterns can be observed.}
\end{figure}

Another possible application for chaotic itinerancy phenomena is as a theoretical
framework for understanding multistability in visual phenomena~\cite{Multistability}. 
Here, an ambiguous figure would correspond to a mixture input where the $Q$ components are the
$Q$ possible interpretations of the figure (see Fig.~\ref{fig7}). 
Visual multistability phenomena will be 
studied computationally and experimentally in another work. A preliminary finding is that,
both in the model and in experiments,
the transition frequency between patterns grows with $Q$, showing that competition between
the $Q$ patterns (and not only fatigue factors) modulates the rate 
of transition between patterns.

Our model differs from previous itinerant systems in regard to:
\begin{itemize}
\item In contrast to Kaneko coupled lattices~\cite{Kaneko},
the number $P$ of quasi-attractors (patterns) can be set in advance and a study in
function of $\alpha=P/N$ can be done;
\item Differently from Nara model~\cite{Nara},
we do not need to dilute the network to obtain itinerancy. This preserves
a large capacity $\alpha_c$ without requiring sophisticated (pseudo-inverse)
learning matrixes.
This large capacity (and larger stability of each individual pattern)
enables us to show that pattern separation (by temporal
segmentation) can be done for inputs with a high level of mixture
(up to $0.6 \alpha$ or higher). 
This should be contrasted to very recent
attempts of using chaotic itinerancy for pattern separation where
mixture inputs have only two or three components \cite{Matykiewicz};
\item In contrast to Aihara \cite{Aihara} and Kaneko \cite{Kaneko} models,
we do not need to use individual chaotic elements to obtain global chaotic 
itinerancy. We think that it is important to stress this point: 
chaotic itinerancy and local chaos are independent concepts;
\item Of course, the advantages have a price: our system is composed of a $O(N^2)$ number of
equations in contrast to $O(N)$ equations of previous itinerant systems.
\end{itemize}

We notice that the same anti-Hebbian process can be implemented
in other rugged landscapes systems like, for example,
a Sherrington-Kirkpatric (SK) model
with zero temperature parallel dynamics.
We expect a similar or more complex itinerant evolution due to the exponential
number of minima in these systems. In such spin glass-like system, however, since
we do not know a priori where the minima lie, we may study energy and
autocorrelation functions instead of the  overlap
$m_\mu(t)$ available in neural networks systems. In these systems, our
anti-Hebbian slow process could be though as an energy paving search method
with some similarity to other methods proposed in the literature \cite{Hansmann}.

A final observation is that we studied only the itinerant
``equilibrium regime'', meaning that in the parameter region
examined the transition frequency matrix appear to have a
time independent form. However, for small step sizes $\epsilon$, one should
expect a scenario where the overall trajectory is transient,
stopping at the most stable attractors.
So, transient itinerant dynamics could be relevant as an alternative
proposal for fast (non thermal) dynamical mechanism to produce convergence
to a ``native'' state in the protein folding problem. It is also conceivable an out
of equilibrium ``glassy'' scenario, probably in the large $\alpha$
regime, where the residence times may strongly depend on the
relative stability of the local minima and the mean residence time
diverges like in trap models \cite{Bouchaud,Martinez2004}.

Out-of-equilibrium chaotic itinerancy is a topic for future work.
These transient and glassy regimes are of interest because the
complex time evolutions of the biological or socio-economical
systems cited in the introduction are primary examples of such
out-of-equilibrium (``historical'') dynamics.  As at least one
concrete suggestion for these problems, our study illustrates the
idea that the competitio (``dialectics'') between short term stabilizing and long
term corrosive \textbf{internal} factors is a sufficient conditions to produce
a punctuated, revolutionary history where, even following a
deterministic dynamics, future is not predictable.

{\textbf Acknowledgments:} Research supported by FAPESP, CNPq and CAPES. We acknowledge
discussions with I. Tsuda, A. Roque da Silva, P. Zambianchi, bibliographic aid from
M. Adachi and K. Aihara and C. F. Schmidt for allowing the use of figure 7.


\begin{thebibliography}{99}

\bibitem{Newman1985} C. M. Newman, J. E. Cohen, and C. Kipnis,
Nature \textbf{315}, 400 (1985).

\bibitem{Drossel} B. Drossel, Adv. Phys \textbf{50}(2), 209 (2001).

\bibitem{Kaneko2003} K. Kaneko and I. Tsuda, Chaos \textbf{13}(3), 926 (2003).

\bibitem{Kaneko} K. Kaneko, Phys. Rev. Lett. \textbf{78}, 2736 (1997).

\bibitem{Tsuda} I. Tsuda, Behav. Brain Sci. \textbf{24}, 793 (2001).

\bibitem{Nara} S. Nara and P. Davis
Prog. Theoret. Phys. \textbf{88}, 845 (1992).

\bibitem{Okuda} H. Okuda and I. Tsuda, Int. J. Bifurcation Chaos \textbf{4},
1011 (1994).

\bibitem{Marcus} C. M. Marcus, F. R. Waugh, and R. M. Westervelt,
Phys. Rev. A \textbf{41}, 3355 (1990).


\bibitem{vanHemmen1997} J. L. van Hemmen, Network: Comput. Neural Syst.
\textbf{8}, V1-V17 (1997).

\bibitem{Anderson1985} A. H. Kawamoto and J. A. Anderson, Acta Psychol.
\textbf{59}, 35 (1985).

\bibitem{Hoshino} O. Hoshino, Y. Kashimori, and T. Kambara.
PNAS textbf{93}, 3303 (1996).

\bibitem{Bouchaud} R. A. Denny, D. R. Reichman, and J.P. Bouchaud
Phys. Rev. Lett. \textbf{90}, 025503 (2003). 

\bibitem{Martinez2004} Martinez A. S., Kinouchi O., and Risau-Gusman S.,
Phys. Rev. E. {\textbf 69}, 017101 (2004).

\bibitem{Freeman} C. H. Skarda and W. Freeman,
Behav. Brain Sci. \textbf{1}, 161 (1987).

\bibitem{Galan} R. F. Galan, M. Weidert, R. Menzel, A. V.  Herz, and C. G. Galizia,
Neural Comput. \textbf{18}, 10 (2006). 

\bibitem{AAC} M. Adachi, K. Aihara, and A. Cichocki,
in International Symposium on Nonlinear Theory and its Application
(NOLTA), Katsurahama-so, Kochi, Japan 93 (1996).

\bibitem{Matykiewicz}  P. Matykiewicz, ICAISC 2004
Lecture Notes in Artificial Intelligence \textbf{3070}, 235 (2004).

\bibitem{Laurent} Laurent, G.
Nat. Rev. Neurosci. \textbf{11}, 884 (2002).

\bibitem{Cleland} T. A. Cleland and C. Linster 
Chem. Senses \textbf{30}, 801 (2005).

\bibitem{smell}R. Gross-Isserof and D. Lancet,
Chem. Senses \textbf{13},191 (1988).

\bibitem{Wright} Wright GA, Thomson MGA, Smith BH 
Proc. Royal Society B - Biol. Sci. \textbf{272}, 2417 (2005). 


\bibitem{Multistability} D. A. Leopold, M. Wilke, A. Maier, and N. K. Logothetis,
Nat. Neurosci. \textbf{5}, 605 (2002).

\bibitem{Multifigure} The figure has been downloaded from 
http://www.rci.rutgers.edu/

\bibitem{Aihara} M. Adachi M. and K. Aihara, Neural Networks
\textbf{10}, 83 (1997).

\bibitem{Hansmann} U. H. E. Hansmann and L. T. Wille, 
Phys. Rev. Lett. \textbf(88), 068105 (2002).







\end{thebibliography}
\end{document}